\documentclass{llncs}

\usepackage{amssymb}
\usepackage{graphicx}
\usepackage{url}
\usepackage{textcomp}

\urldef{\mails}\path|{hzgirgis, bharat}@cse.buffalo.edu|

\begin{document}

\mainmatter  

\title{JavaTA:\\ A Logic-based Debugger for Java}


%
%
\author{ Hani Z. Girgis \and Bharat Jayaraman}

\institute{Department of Computer Science and Engineering\\
University at Buffalo, The State University of New York\\
Buffalo, NY 14260, USA\\
\mails\\
\url{http://www.cse.buffalo.edu}}

%
%

\maketitle

\begin{abstract}
This paper presents a logic based approach to debugging Java programs. In contrast with traditional debugging we propose a debugging methodology for Java programs using logical queries on individual execution states and also over the history of execution. These queries were arrived at by a systematic study of errors in object-oriented programs in our earlier research.  We represent the salient events during the execution of a Java program by a logic database, and implement the queries as logic programs. Such an approach allows us to answer a number of useful and interesting queries about a Java program, such as the calling sequence that results in a certain outcome, the state of an object at a particular execution point, etc. Our system also provides the ability to compose new queries during a debugging session. We believe that logic programming offers a significant contribution to the art of object-oriented programs debugging.

\end{abstract}
\section{Introduction}
This paper shows some of the benefits of applying logic programming techniques in the debugging of object-oriented programs.  Debugging object-oriented programs has traditionally been a procedural process in that the programmer has to proceed step-by-step and object-by-object in order to uncover the cause of an error. In this paper, we propose a logic-based approach to the debugging of object-oriented programs in which debugging data can be collected via higher level logical queries. We represent the salient events during the execution of a Java program by a logic database, and implement these queries as logic programs. Such an approach allows us to answer a number of useful and interesting queries about a Java program.

To illustrate our approach, note that a crucial aspect of program understanding is observing how variables take on different values during execution.  The use of print statements is the standard procedural way of eliciting this information.  This is a classic case of the need to query over execution history. Other examples include queries to find which variable has a certain value; the calling sequence that results in a certain outcome; whether a certain statement was executed; etc. We arrived at a set of queries by a study of the types of errors that arise in object-oriented programs \cite{hzgirgis:b}. 

We propose two broad categories of queries in this paper: (i) queries over individual execution states and (ii) queries over the entire history of execution, or a subset of the history. Our proposed method recognizes the need to query sub-histories; such capability is especially useful when debugging large scale software whose program trace is composed of millions of execution events.  Our system has the ability to filter system objects so that a programmer may focus on the objects explicitly instantiated form user defined classes.
  
Our current implementation, called JavaTA, takes a Java program as input and builds a logic database of salient events (method call, return, assignment, object creation, etc) during the execution of a Java program using the JPDA interface (Java Platform Debugger Architecture). Our approach to recording the history of changes is incremental in nature, i.e., when a variable is assigned, we save only the new value assigned to the variable. Thus, queries about previous execution states involve some state reconstruction.  A textual interface allows the user to pose a number of queries as detailed in section \ref{QueriesonProgramTrace}.  
  
Thus the contributions of our paper are: (1) logic-based approach to debugging object-oriented programs; (2) the provision of queries over individual states and the history of execution; (3) a prototype for a trace analysis for object-oriented programs.

The remainder of the paper is organized as follows. Section \ref{OverviewofLogic-basedDebugging} presents an example, called the 'traveling null pointer', in order to illustate our overall approach. Section \ref{JavaTAArchitecture} presents the architecture of JavaTA, along with the Java Event Log language. Section \ref{QueriesonProgramTrace} outlines the principles of our debugging methodology. Section \ref{relatedWork} surveys closely related research and compares them with our work. Section \ref{conclusions} presents conclusions and areas of further research.
%
\section{Overview of Logic-based Debugging}
\label{OverviewofLogic-basedDebugging}
This section provides an overview of our approach to logic-based debugging with an example.  We present the 'traveling null pointer' example, which illustrates a bug pattern in which a method call incorrectly returns a null pointer and the client of that method propagates the null pointer through a call chain, and, finally, a null pointer exception is thrown when the client code of the last call in the chain tries to de-reference the null pointer. In other words, the code that originates the null pointer and the code that de-references that pointer are far apart spatially and temporally. Fig. \ref{fig:TheTravelingNullPointerDefectPattern} illustrates the traveling null pointer defect pattern in Java code. The instance method {\tt doSomeThing} in {\tt FarAWayClass} returns a null pointer due to erroneous conditions.  When this program is executed it reports a null pointer exception at line 14.

\begin{figure}
\centering
\begin{scriptsize}
\begin{verbatim}
public class Example {
  public Example() {
    m1();
  }
  public void m1() {
    FarAWayClass o = new FarAWayClass();
    String result = o.doSomeThing();
    m2(result);
  }
  public void m2(String result) {
    mN(result);
  }
  public void mN(String result) {
/*14*/if (result.equals("some result"))
        System.out.println("some result");
      else
        System.out.println("other result");
  }
  public static void main(String[] args) {
    new Example();
  }
}
class FarAWayClass {
  public String doSomeThing() {
    // some code that incorrectly results in returning a null
    return null;
  }
}
\end{verbatim}
\end{scriptsize}
\caption{The traveling null pointer bug pattern}
\label{fig:TheTravelingNullPointerDefectPattern}
\end{figure}

JavaTA generates a trace for the example program.  (We use the term 'trace' and 'execution history' interchangeably in this paper.)  The trace includes 17 events as shown in Fig. \ref{fig:TheExampleProgramTrace}. JavaTA recorded thread start, thread exit, method call, method exit, exception, and step events. Trace events are descried in JEL, a Prolog-based description language for program trace. For example, the second event recorded has a unique id 1 and it belongs to the main thread. The event has been recorded due to the invocation of a method called main. The term {\tt l('Example.java', 20)} indicates that the method is defined in the Example.java file on line 20. The term {\tt c('Example')} means that the method is a class or static method of the Example class. The main method takes an instance of array of strings as the only argument. Instance or object is described by the class name and a unique id as in the term {\tt o('java.lang.String[]', 641)}. 

\begin{figure}
\begin{scriptsize}
\begin{verbatim}
event(0, 'main', threadstart('main')).
event(1, 'main', methodcall(l('Example.java', 20), c('Example'), 'main', 
                              [o('java.lang.String[]',641)])).
event(2, 'main', methodcall(l('Example.java', 2), o('Example', 643), '<init>',[])).
event(3, 'main', step(l('Example.java', 3), [])).
event(4, 'main', methodcall(l('Example.java', 6), o('Example', 643), 'm1', [])).
event(5, 'main', methodcall(l('Example.java', 23), o('FarAWayClass', 645), '<init>', [])).
event(6, 'main', methodexit(5, l('Example.java', 23), o('FarAWayClass',645),'<init>','void')).
event(7, 'main', step(l('Example.java', 6), [])).
event(8, 'main', step(l('Example.java', 7), [lv('o', o('FarAWayClass', 645))])).
event(9, 'main', methodcall(l('Example.java', 26), o('FarAWayClass',645),'doSomeThing',[])).
event(10, 'main', methodexit(9, l('Example.java', 26), o('FarAWayClass', 645), 
                  'doSomeThing', 'null')).
event(11, 'main', step(l('Example.java', 7), [lv('o', o('FarAWayClass', 645))])).
event(12, 'main', step(l('Example.java', 8), [lv('o', o('FarAWayClass' ,645)), 
                       lv('result', 'null')])).
event(13, 'main', methodcall(l('Example.java', 11), o('Example', 643), 'm2',['null'])).
event(14, 'main', methodcall(l('Example.java', 14), o('Example', 643), 'mN',['null'])).
event(15, 'main', exception(l('Example.java', 14), o('java.lang.NullPointerException', 666), 
                              'null', uncaught)).
event(16, 'main', threaddeath('main')).
\end{verbatim}
\end{scriptsize}
\caption{Program trace for Traveling Null Pointer Example}
\label{fig:TheExampleProgramTrace}
\end{figure}

To facilitate trace analysis JavaTA provides a set of predefined queries. Table \ref{tab:IntrogationSession} shows the three predefined queries used in the debugging session. First the user asked regarding the environment where the exception is thrown as in Q1. A1 indicates that the enclosing method is mN whose single argument is null and the call to the enclosing method occurred at event id 14. The current question is where the null pointer originated. Q2 inquires full detail call chain leading to event id 14. A2 shows that method m1 called method m2 which called method mN. The initial call to method main and the constructor is omitted for simplicity of presentation. By investigating the argument passed to m2 it is clear that it has a null value. Method m2 is called from m1, and m1 is called at event id 4. When looking at the source code of method m1, the programmer concludes that the local variable `result' holds a null value since it is passed as the argument to m2. At this point all fingers point to the method o.doSomeThing as the cause of the error. To verify that o.doSomeThing returns null pointer, Q3 asks for the details of all methods that were called before m2 in the same enclosing environment. A3 confirms that o.doSomeThing has returned the null value. 	

\begin{table}
	\caption{Debugging session}
	\label{tab:IntrogationSession}
	\centering
	\begin{scriptsize}
	\begin{tabular}{ll}
		\hline
		\hline
		Q1&?-where\_exception\_is\_thrown('main', Environment).\\                                                    
		A1&Environment = event(14,main,methodcall(l('Example.java',14),\textbf{o('Example',643),mN},[null]))\\
		\hline
		Q2&?-full\_detail\_call\_chain( 15 , CallChain ).\\
		A2&[event(14,main,methodcall(l(Example.java,14),o(Example,643),\textbf{mN,[null]})),\\
		  & event(13,main,methodcall(l(Example.java,11),o(Example,643),\textbf{m2,[null]})),\\
		  & event(4,main,methodcall(l(Example.java,6),o(Example,643),\textbf{m1},[])), ..]\\
		\hline
		Q3&?-pre\_event\_called\_methods(13 , PreCalled).\\
	  A3&[\textit{[}event(5,main,methodcall(l(Example.java,23),o(FarAWayClass,645),\textlangle init\textrangle,[])),\\	   
	    &event(6,main,methodexit(5,l(Example.java,23),o(FarAWayClass,645),\textlangle init\textrangle,void))\textit{],}\\      			
      &\textit{[}event(9,main,methodcall(l(Example.java,26),o(FarAWayClass,645),\textbf{doSomeThing},[])),\\    
	    &event(10,main,methodexit(9,l(Example.java,26),o(FarAWayClass,645),\textbf{doSomeThing,null}))\textit{]}]\\ 
		\hline
	\end{tabular}
	\end{scriptsize}
\end{table}

The Prolog code for the three queries referenced in table \ref{tab:IntrogationSession} are shown in section \ref{QueriesonProgramTrace}. Given that these are frequently used queries in object-oriented program debugging and also noting that the average Java programmer may be unfamiliar with Prolog, JavaTA provides these queries as built-in primitives. Several additional useful debugging queries and their Prolog implementation are also illustrated in   section \ref{QueriesonProgramTrace}.
%
\section{JavaTA Architecture}
\label{JavaTAArchitecture}
We have implemented a prototype of the JavaTA framework as a distributed system. Fig. 3 shows the main tiers and components of the framework. The architecture of JavaTA is composed of four tiers. The first tier consists of three components: the JPDA, the Prolog server, and the built-in primitives. JPDA, the Java Platform Debugger Architecture \cite{jpda}, is designed as a distributed system that can interface with a JVM running on the same machine or on a different machine. Prolog Beans \cite{prolog:beans} is a Prolog server that can be interfaced with Java or .Net. The client-server architecture of Prolog Beans allows the server to be a component of a distributed system. Prolog Beans was designed to handle large applications.

The second tier is composed of two components: the Logger and the Query Manager. Once the Logger receives a Java program it starts a JVM and subscribes for the desired events with the JPDA. It is also possible (but not implemented in the current prototype) that the Logger interacts with an already running JVM. The Query Manager is responsible for constructing Prolog goals and sending the constructed goals to the Prolog Beans server. Once the Query Manger receives answers, it forwards them back to the Tools Interface. The third tier is composed of only one component: the Tools Interface which is a facade for the JavaTA Framework. The fourth tier has only one component: the User Interface that interacts with the Tools interface and the user.

\begin{figure}
	\centering
		\includegraphics[width=12cm]{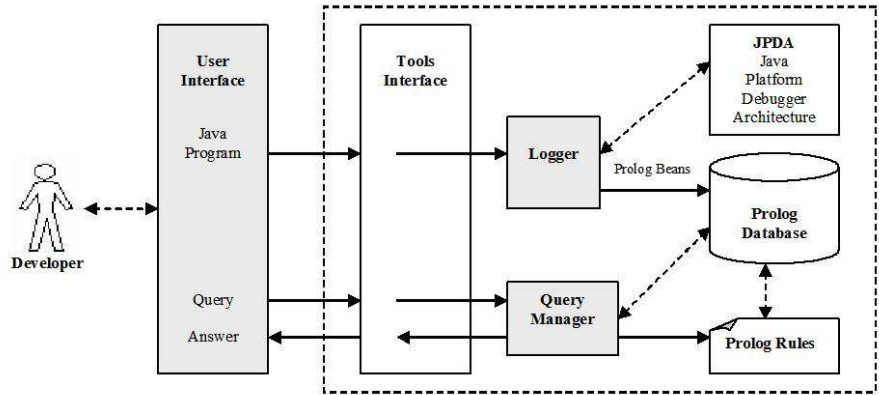}
	\caption{JavaTA architecture}
	\label{fig:JavaTA-architecture}
\end{figure}

\textbf{JEL}(Java Events Log) is a program trace description language. JIVE \cite{pvg:a,pvg:b} and JyLog \cite{jylog} have implemented similar recording techniques based on logging in XML; however, JEL describes the program trace as a set of Prolog facts. JEL can be easily extended to include a sophisticated description of static and dynamic information about a given program. Table \ref{tab:PartOfJELBNF} shows part of the BNF grammar of JEL. The basic construct in JEL is the event term. Each event has a unique id and thread in addition to other specific information. Objects are identified by their class and a unique id. The implemented prototype supports the description of the following nine events.
 
\begin{enumerate}
	\item 	Method call event records the source code location of the first executable line of the method body, the class or the instance that this method was invoked on, method name, and method arguments. 
	\item 	Method exit event records similar information as method call event, in addition to the id of the corresponding method entry event and the returned value instead of the arguments. 
	\item 	Set Field event records the source location where the field was set to a new value, the instance or the class where this field is declared, and the new value. 
	\item 	Data Structure event is recorded after a method entry event, method exit event, and set field event if the type of the field being assigned a new value is a data structure. The data structure can be an array or a Collection instance. The event describes the source code information of the event that caused the recording of the data structure. 
	\item 	Step event describes the source code location in addition to names and values of visible local variables in each step. Each step corresponds to the execution of a source line.  
	\item 	Exception event records the source code location, the exception instance, the exception message, and the catch location if it is caught or the uncaught keyword other wise.
	\item 	Thread Start and Thread Death events record the starting or the ending of a thread. The thread group is also recorded.
	\item 	Member fields event records information regarding member fields of a given class.
\end{enumerate}

\begin{table}
	\caption{Part of JEL BNF}
	\label{tab:PartOfJELBNF}
	\centering
	\begin{scriptsize}
	\begin{tabular}{ll}
	\hline
	\hline
	\textlangle events\textrangle &::= event*\\
 	\textlangle event\textrangle &::= event '(' \textlangle id\textrangle, \textlangle thread\textrangle, \textlangle execution-event\textrangle  ')' '.'\\
 	\textlangle execution-event\textrangle &::= \textlangle member-fields\textrangle \textbrokenbar \textlangle method-call\textrangle  			     
\textbrokenbar \textlangle method-exit\textrangle \textbrokenbar \textlangle set-field\textrangle \textbrokenbar \textlangle data-structure\textrangle  \textbrokenbar\\   
 &\textlangle exception\textrangle \textbrokenbar \textlangle step\textrangle \textbrokenbar  \textlangle thread-start\textrangle \textbrokenbar \textlangle thread-death\textrangle \\ 
 	\textlangle method-call\textrangle &::= methodcall '(' \textlangle location\textrangle, (\textlangle instance 	\textrangle \textbrokenbar \textlangle class\textrangle ),  \textlangle name\textrangle ,  \textlangle arguments\textrangle  ')'\\
	\textlangle method-exit\textrangle &::= methodexit '(' \textlangle id\textrangle, \textlangle location\textrangle , (\textlangle instance 	\textrangle  \textbrokenbar \textlangle class\textrangle ),  \textlangle name\textrangle ,  \textlangle value\textrangle ')'\\
 	\textlangle set-field\textrangle &::= setfield '('  \textlangle location\textrangle , ( \textlangle instance \textrangle  \textbrokenbar   \textlangle class\textrangle ),  \textlangle name\textrangle ,  \textlangle value\textrangle ')'\\
 	\textlangle data-structure\textrangle &::= datastructure'('  \textlangle location\textrangle, \textlangle contents\textrangle ')'\\
 	\textlangle exception\textrangle &::= exception '(' \textlangle location\textrangle, \textlangle instance\textrangle, \textlangle message\textrangle, (\textlangle location\textrangle \textbrokenbar uncaught) ')'\\
 	\textlangle step\textrangle &::= step '(' \textlangle location\textrangle, \textlangle local-variable-list\textrangle  ')'\\
 	\textlangle member-fields \textrangle &::= memberfields '(' \textlangle class\textrangle  ,  \textlangle member-fields\textrangle  ')'\\
 \textlangle thread-start\textrangle &::= threadstart '(' \textlangle thread-group\textrangle   ')'\\
 \textlangle thread-death\textrangle &::= threaddeath '(' \textlangle thread-group\textrangle   ')'\\
 \hline
	\end{tabular}
	\end{scriptsize}
\end{table}
\section{Queries on Program Trace}
\label{QueriesonProgramTrace}
The debugging process involves three phases: (i) formulating a hypothesis about the root of the error; (ii) collecting program-specific data that is pertinent to the hypothesis; (iii) analyzing the collected data to prove or disprove the hypothesis. The difference between JavaTA  and traditional debugging lies in their respective approaches to the data collection phase (ii).  In JavaTA, data collection is performed by high-level queries on the trace. In traditional debugging, data collection is performed by the programmer by a process of manually stepping through the code, setting break points, and inspecting objects.

In this section, the program trace is recorded as a Prolog database.  The database is populated by entries corresponding to execution events which are specified by JEL.  While it is possible to pre-process this database in order to construct auxiliary structures such as call trees, we do not resort to such optimizations here, but present a relatively straightforward implementation of the debugging primitives directly in terms of the event database.  The debugging primitives, or predefined queries, provided by JavaTA can be organized under three categories: queries on specific events, queries on execution history, and query management.  Section \ref{subsec:QueriesOnProgramState} discusses queries on specific events.  There are four kinds of queries over the execution history and are illustrated in section \ref{QueriesOverExecutionHistory}. Query management and programmability techniques are discussed in section \ref{QueryManagement}.

\subsection{Queries on Program State}
\label{subsec:QueriesOnProgramState}
\textbf{Group Method Calls According to Call Chain.} Compared with the traditional procedural paradigm, the object-oriented paradigm engenders the use of many small methods and greater method interaction. Thus, posing queries regarding the interaction between objects is essential in the debugging process and in the understanding of object oriented programs in general. A method call can be viewed as a message whose content is the passed arguments. Each message has a response which is the returned value or void. A message can have no response if it exits abnormally, i.e. throws an exception. Call chain can serve as a way to know the execution path leading to the execution of a specific event or as a way to inspect argument values propagated through the chain of calls. Fig. \ref{fig:TheCallChainRuleInProlog} illustrate the \textit{call\_chain} rule in Prolog.

\begin{figure}
\centering
\begin{small}
\begin{verbatim}
call_chain(Id, OutList):-
  findall(CId, any_enclosing_method(Id, CId ,_), OutList).

% normal termination
any_enclosing_method(Id, CallId, ExitId):-
  event(Id, T , _),
  event(CallId, T, methodcall( _ ,O , N , _)),
  Id > CallId,
  event(ExitId, T, methodexit(CallId, _, O, N, _)),
  Id < ExitId.

% abnormal termination
any_enclosing_method(Id, CallId, ExpId) :-
  event(Id, T, _),
  event(CallId, T, methodcall(_, O, N, _)),
  Id > CallId,
  \+ event(_, T, methodexit(CallId, _, O, N, _)),   
  event(ExpId, T, exception(_, _, _, uncaught)),
  Id =<  ExpId.
\end{verbatim}
\end{small}
\caption{The call chain rule}
\label{fig:TheCallChainRuleInProlog}
\end{figure}

The rule \textit{any\_enclosing\_method} specifies any enclosing method for a given event. For example, suppose method $m1$ called method $m2$ where event $e$ was executed. Let \ensuremath{id_{m1}^c, id_{m1}^e, id_{m2}^c, id_{m2}^e, id_{e}} are the id's for the following events: $m1$ call, $m1$ exit, $m2$ call, $m2$ exit, and the execution of event $e$ respectively, assuming that the program has terminated normally. Note that \ensuremath{id _{m1}^c < id _{m2}^c < id _{e} <  id _{m1}^e < id _{m2}^e}. Event $e$ is enclosed in method $m2$ which is enclosed in method $m1$; therefore, methods $m1$ and $m2$ are considered as enclosing methods. According to the rule \textit{any\_enclosing\_method} \ensuremath{CallId = id _{m1}^c , ExitId =  id _{m1}^e} or \ensuremath{CallId = id _{m2}^c, ExitId =  id _{m2}^e}. Therefore, a call chain leading to the execution of a given event is all the enclosing methods for that event. According to the \textit{call\_chain} rule, OutList = [\ensuremath{id_{m1}^c, id_{m2}^c}].   

\textbf{Query Where an Event Occurred.} In object-oriented programming, execution events occur within an environment. An environment is an instance object and an instance method invocation or else it is a class and a static method invocation. This environment represents the enclosing environment for an event. The instance or the class is referred to as the enclosing instance or enclosing class and the method is referred to as the enclosing method for the event. The enclosing environment for the following four events: (1) set field event, (2)  method call event, (3) object instantiation event which is recorded as method call event to the method \textlangle init\textrangle, and (4) exception event can be obtained by the where query. 

Fig. \ref{fig:TheWhereRuleInProlog} shows the rule \textit{where\_exception\_is\_thrown} for a given thread. Once an exception is thrown, the thread in which the exception occurred is terminated; therefore, there is at most one uncaught exception per thread. The rule \textit{where} specifies that the enclosing environment for a given event is the first call in the reversed call chain specified by the \textit{full\_detail\_call\_chain} rule which reverse the list of id's obtained form \textit{call\_chain} rule and extract the associated events from the database. 

\begin{figure}
\centering
\begin{small}
\begin{verbatim}
where_exception_is_thrown(Thread, Environment):-
  where(Thread,exception(_, _ ,_ , uncaught), Environment).

where(Thread, Event, EnclosingEnvironment):-
  event(Id, Thread, Event),  
  full_detail_call_chain(Id, [EnclosingEnvironment|_]).
\end{verbatim}
\end{small}
\caption{The where rule}
\label{fig:TheWhereRuleInProlog}
\end{figure}

\textbf{Query the State of an Object.}  Querying the state of an object is concerned with the encapsulation aspect of object-oriented programming. The state of an object is captured in the values of its member fields and public and protected member fields of its super classes. The rule \textit{object\_state} in Fig. \ref{fig:TheStateRuleInProlog} illustrates how the state of the object OName whose id is OId can be reconstructed at event id E. Object instantiation event is recorded as a method call to \textlangle init\textrangle. The domain of the \textit{object\_state} rule is the segment of the program history between id S when the instantiation occurred and id E which is specified by the user. Member fields of a class is recorded as memberfields event. The \textit{object\_state\_helper} rule specifies field value contributing to the desired state as the last value in the field history between id S and id E. Rule \textit{instance\_field\_history} is discussed in the next section. 

For example, let \ensuremath{id_{init} , id_{end}} be the boundary of the search domain and \ensuremath{f_1 , f_2} are fields of the desired object. Suppose that histories of fields \ensuremath{f_1 , f_2} are \ensuremath{\{\{id_{i} , f_1, v_{i}\},..,\{id_{j},f_1 , v_{j}\}\}} and \ensuremath{\{\{id_{n} , f_2,v_{n}\},..,\{id_{m} ,f_2, v_{m}\}\}} respectively, $v_k$ and $id_k$ stand for the value of the field and when it was assigned respectively. Note that \ensuremath{id_{init} < id_{i}, id_{j}, id_{n} , } \ensuremath{id_{m} <=id_{end}}, and \ensuremath{ id_{i} < id_{j}}, and \ensuremath{id_{n} < id_{m}}. Then the object's state is \ensuremath{\{\{id_{j} ,f_1, v_{j}\} \{id_{m} ,f_2, v_{m}\}\}}.

\begin{figure}
\centering
\begin{small}
\begin{verbatim}
object_state(E, OName, OId, State):-
  event(S, _, methodcall(_, o(OName, OId), '<init>', _)),
  event(_, _, memberfields(c(OName), Fields)),
  object_state_helper(S, E, OName, OId, Fields, [], State).

object_state_helper(_, _, _, _, [], Ans, Ans).

object_state_helper(S, E, OName, OId, [cf(F)|R], SoFar, Ans):-
  class_field_history(S, E, OName , F, H),
  last(H, C),
  object_state_helper(S, E, OName, OId, R, [C|SoFar], Ans).

object_state_helper(S, E, OName, OId, [of(F)|R], SoFar, Ans):-
  instance_field_history(S, E, OName, OId ,F, H),
  last(H, C),
  object_state_helper(S, E, OName, OId, R, [C|SoFar], Ans).
\end{verbatim}
\end{small}
\caption{The state rule}
\label{fig:TheStateRuleInProlog}
\end{figure}

\textbf{Queries On Method State} In design-by-contract (DBC) \cite{meyer:a,meyer:b,mitchell} the client has to meet preconditions or specific requirements in order to be able to call a certain method. These requirements are usually constraints on the arguments and the state. Our method generalizes the requirement to be imposed on any execution event and not only on method calls as in DBC. The following three factors can affect the execution of a given event within the enclosing method: (1) arguments values, (2) the returned value of all preceding method calls to a given event within the same enclosing method. Fig. \ref{fig:ThePreEventCalledMethodsInProlog} shows the \textit{pre\_event\_called\_methods} rule, and (3) local variables values before the execution of the event. Thus those three factors are considered candidate queries.

Analogously, the post-condition in DBC is the effect that the called method promises upon its correct completion. Our methodology generalizes this idea to all executed events. The effect of the execution of an event on the enclosing method can appear in the following three areas: (1) the returned value of the enclosing method, (2) methods that have been called after the execution of the event within the same enclosing method, and (3) Local variables values after the execution of the event. DBC is not capable of specifying directly that some other methods need to be called before or after a given method. Having recorded the execution history it is possible to inspect whether a certain method(s) has been called before or after a given event.

\begin{figure}
\centering
\begin{small}
\begin{verbatim}
% all methods which are called and exited before Id 
% and within the same enclosing method
pre_event_called_methods(Id, OutList):-
  findall([Call, Exit], pre_event_called_method(Id, Call, Exit), OutList).
	
pre_event_called_method(Id, event(Call, T, E1), event(Exit, T, E2)):-
  enclosed_method(Id, Call, Exit),
  Call < Id, Exit < Id,
  event(Call, T, E1), event(Exit, T, E2).

% any method which is called and exited within 
% the enclosing method of event whose id is Id 	
enclosed_method(Id, Call, Exit):-
  full_detail_call_chain(Id, [event(EnclosingId, Thread, _)|_]),
  any_enclosing_method(Id, EnclosingId, IdExit),
  event(Call, Thread, methodcall(_, O, N, _)),
  event(Exit, Thread, methodexit(Call, _, O, N, _)),
  any_enclosing_method(Call, EnclosingId, IdExit),
  any_enclosing_method(Exit, EnclosingId, IdExit).
\end{verbatim}
\end{small}
\caption{The pre-event called methods rule}
\label{fig:ThePreEventCalledMethodsInProlog}
\end{figure}

\subsection{Queries over Execution History}
\label{QueriesOverExecutionHistory}
\textbf{Execution History Subset.} The programmer should have the ability to focus on an interval of the execution history when an erroneous behavior is suspected to occur. Such feature is useful in dealing with large program trace by allowing the programmer to filter out irrelevant data.

\textbf{Gathering Data.} Eisenstadt \cite{eisenstadt} in his study on how bugs were found in 51 cases gathered from professional programmers found that 
programmers have used the following 4 techniques to locate the defect root: data gathering, code inspection, expert help, and controlled experiments. In 27 cases the bugs were found by gathering data regarding the execution of the program. JavaTA can gather data automatically regarding the following (1) member field value history; (2) local variable value history which is important in understating loop execution; (3) history of arguments of method calls; (4) history of return value of method calls; (5) history of contents of data structure; (6) all class instances and their states which is important in understanding user defined data structures; (7) thread status such: running and exited threads. 

Fig. \ref{fig:TheInstanceFieldHistoryInProlog} shows the rule for \textit{instance\_field\_history}. The rule specifies a segment of the history between id S and id E for an instance field F of object OName whose unique id is OId. The rule \textit{instance\_field\_value} specifies that a value of a given field can be obtained from a set field event provided that its id is between S and E.

\begin{figure}
\centering
\begin{small}
\begin{verbatim}
instance_field_history(S, E, OName, OId, F, R):-
  findall([Id, F, V], instance_field_value(S, E, OName, OId, F, Id, V), R).

instance_field_value(S, E, OName, OId, F, Id, V):-
  event(Id, _ , setfield(_ , o(OName ,OId), F, V)), 
  Id >= S, Id =< E.
\end{verbatim}
\end{small}
\caption{The instance field history}
\label{fig:TheInstanceFieldHistoryInProlog}
\end{figure}
     
\textbf{Call Tree.} Grouping method calls according to a call tree is motivated by the need to depict interactions among objects. Call tree can be defined as methods called by the method of interest. Method calls that are involved in a call tree collaborate in achieving one task. Those methods are not necessarily dependent on each other, unlike method calls in a call chain in which the called method depends on the caller.   

\textbf{Query about Statement Execution.} One of the most recurring questions in the debugging process is whether a certain statement has been executed or not. Novice programmers find the answer for such a question by inserting multiple print statements in program's code. Advanced developer would insert break points using a traditional debugger to verify whether a given statement has been executed or not. The answer to this question is either yes or no.  We propose the following seven queries. (1) Was a given conditional statement executed? (2) Was a given method called? (3) Was a member field assigned to a given value? (4) Is there an instance of a specific class? (5) Was a specific exception caught? (6) Is a given thread still running? (7) Has a given thread exited?

\subsection{Programmability and Query Management}
\label{QueryManagement}
\textbf{Compose and Save Queries.} The ability to compose queries provides a way to adapt queries to recurring bug patterns as well as to the individual needs of the developer. The idea is similar to the idea behind the Emacs system that allows the user to add macros dynamically to add functionality to the system. Composed queries guarantee the flexibility and the extendibility of our framework. Allowing the user to add queries dynamically results in a general purpose static analyzer for program trace. However, we do not have experimental data to support our claim especially on large program traces or for more complicated analyses. 

Liang and Kai \cite{liang} developed a scenario-driven debugger. The idea is to allow the programmer to model a behavior view for a specific task as finite automata. The debugger allows the programmer to inspect the task execution progress. A similar capability can be added to JavaTA by composing a Prolog rule. Fig. \ref{fig:SpecificationOfTheBehaviorViewForTheLoginTaskInProlog} shows the \textit{login} Prolog rule used to inspect the execution of the login task. The original example of the login task and its behavior view is illustrated in Liang and Kai's paper \cite{liang}. A standard login task is composed of (i) obtaining the user name (ii) obtaining the password (iii) verifying the user name and the password. If any step fails the login process fails, otherwise the user is allowed to login. One important difference is the analysis used in JavaTA is postmortem analysis. On the other hand, the scenario-driven debugger uses on-line analysis.

\begin{figure}
\centering
\begin{small}
\begin{verbatim}
login :-
  event(_, _, setfield(_, _, 'uBox', UBox)),
  event(_, _, setfield(_, _, 'pBox', PBox)),
  event(Id1, _, methodexit(_, _, UBox, 'getText', Username)),
  write('got user name'), nl,
  event(Id2, _, methodexit(_, _, PBox, 'getText', Password)),
  Id2 > Id1, write('got password'), nl,
  event(Id3, _, methodcall(_, _ , 'verify', [Username, Password])),
  Id3 > Id2, write('enter verify'), nl,
  event(Id4, _, methodexit(Id3, _, _, 'verify', 'true')),
  Id4 > Id3, write('exit verify'), nl,
  event(Id5, _, methodexit(_, _, _, 'login', 'true')),
  Id5 > Id4, write('exit login'), nl.
\end{verbatim}
\end{small}
\caption{Specification of the behavior view for the login task}
\label{fig:SpecificationOfTheBehaviorViewForTheLoginTaskInProlog}
\end{figure}

\textbf{Comparing Query Results.} Eisenstadt \cite{eisenstadt} describes the "Dump \& Diff" as a technique to locate errors. This technique works as follows. The output of print statements is saved to two text files corresponding to two different executions; the two files are then compared using a source-compare "diff" utility, which highlights the difference between the two outputs. This technique can be adapted to query multiple execution histories and to compare the results of multiple queries over the same execution history. Comparative queries can be helpful to see the difference between data structure contents, and call chains and much more. Comparative queries can also be applied to isolating errors related to software maintenance by posing a query on two runs obtained form two versions and comparing query results.

\textbf{Save Queries Answers.} Calculating a query on a large program history is costly and time demanding. In many debugging scenarios the programmer may go back to examine the results of previous queries or would like to compare them. Re-computing a query on such execution history is wasteful; therefore, queries and their answers should be saved. The WhyLine \cite{ko} allows for data provisioning to ease the debugging process; however, JavaTA adapt this technique due to the cost associated with query evaluation on large program trace. 
%
\section{Related Work}
\label{relatedWork}     

Several research projects apply the concept of program trace analysis to program understanding and debugging. This section starts by giving a brief overview of each system, and then a comparison between the JavaTA and the related projects according to four criteria: (1) automatic program trace extraction; (2) program trace navigation features; (3) query language support; (4) built-in trace analyses. 

Opium \cite{ducasse:a,ducasse:b} is a Prolog-based program trace analysis language for Prolog. The language has several built-in primitive constructs to allow the navigation of the program trace. The trace may be extracted through interaction with a Prolog interpreter. Opium uses on-line dynamic trace analysis, in addition to a trace database if needed. Opium provides a set of built-in abstract views of Prolog execution. This set is based on cognitive study on how programmers understand functionality. Several extensions to Opium have been developed to perform loop analysis, failure analysis, dynamic slicing, profiling and other analyses.

Ducass\'{e} has proposed a trace-based debugger for C called Coca \cite{ducasse:c}. The trace query language is Prolog with eight built-in primitive predicates. Coca is built on top of the GDB. In order to be able to extract program trace from the GDB, Coca transforms the source code to be able to use the breakpoint mechanism in the GDB. Once a user enters a query, Coca computes a set of source lines that may contribute to the answer, and then instructs GDB to set breakpoints on these lines. Coca uses dynamic trace analysis and does not build a database for the program trace.   

Lencevicius et al \cite{lencevicius} proposed a query-based debugger to understand object relationships.  Their query language is expressed in the same language as the target object-oriented language (Self), and thus a programmer does not need to learn a new language. Queries consist of a search domain and a constraint. Lencevicius' query-based debugger provides incremental delivery of results, a feature that is useful in dealing with queries that takes considerable time to find all answers.  

Recently, PQL (Program Query Language) was developed by Martin et al \cite{martin} to query over source code and program trace for finding errors and security flaws in programs.  Queries may formulate application-specific code patterns that may result in vulnerabilities at run-time. Queries are translated to Datalog (which is essentially declarative Prolog without function symbols), and provide the ability to take an action once a match found. A combination of static and dynamic analyses is performed to answer queries.  The PQL compiler generates code that is weaved into the target application and matches against a history of relevant events at execution time. A number of interesting security violations are found by this technique.

Goldsmith et al \cite{goldsmith} proposed the PTQL (Program Trace Query Language) as a relational query language designed to query program trace.  Similar in goals with PQL, PTQL employs an SQL-like query language. Partiqle compiles the PTQL queries into instrumentation in a given Java program.  PTQL queries can be used to specify what is to be recorded during program execution, and hence this technique can be effective with programs that generate many irrelevant events.

Hy$^+$ \cite{consens} is a visual debugger for distributed programs. The system works as follows. The program is instrumented to obtain trace which is used to build database implemented in CORAL. Programmers can specify debugging queries and visualizations using a visual declarative query language called GraphLog. These visual queries once formulated can be saved and applied to other programs since these queries are application independent. This technique allows the programmere to visualize a specific program behavior pattern and filter out irrelevant events. Hy$^+$ performs static trace analysis and has a simple postmortem dynamic trace analysis by animating the program trace.     

JIVE's \cite{pvg:a,pvg:b} (Java Interactive Visualization Engine) design is based on the following seven criteria: (1) depict objects as environment of method execution; (2) display object states at different levels of granularity; (3) provide a sequence diagram to capture the history of execution; (4) support forwards and backwards execution of programs; (5) support queries on the runtime state; (6) produce clear and legible drawings; (7) uses exiting Java technologies. JIVE interacts with the JPDA to extract program trace. An on-line dynamic trace analysis is applied while the program runs for the first time in the forwards direction and postmortem trace analysis is applied in the backwards direction or in the forwards direction once program terminates. 

The omniscient debugger (ODB) developed by Bil Lewis \cite{lewis} aims at easing the navigation of program trace in both forwards and backwards directions. ODB obtains program trace by a load-time instrumentation of the byte code of the debugged program. Execution events are recorded while the program runs, once finished a program state display is provided. ODB uses a static trace analysis and the program trace is kept in memory. Lewis proposed three techniques to reduce the size of the recorded program trace (1) delete old events; (2) allow the programmer to exclude a set of classes and methods form instrumentation and recording 3) a recording interval can be specified. The recording technique applied in the ODB is fast and efficient. 

WhyLine \cite{ko} is an interrogative debugger for the Alice programming environment. It allows the user to ask why a given event did or did not occur. The WhyLine gives the answer in the form of an execution path that leads or was supposed to lead to the execution of the given event. The path is annotated with control flow information.
 
The comparison among these systems are based on four features: (1) automatic program trace extraction; (2) program trace navigation features such as forwards and backwards stepping, breakpoint, and conditional breakpoint; (3) query language support; (4) built-in trace analyses including a set of the most recurring debugging queries or abstract views of program behavior. Table \ref{tab:ToolsComparison} shows the comparison among the 10 system. JavaTA currently does not support trace navigation; however, it is straight forward to implement. Opium supports similar features as JavaTA especially the built-in trace analyses; however, these analyses are hard to be compared since they target two different programming paradigms namely declarative and imperative respectively.

\begin{table}
	\caption{Tools Comparison}
	\label{tab:ToolsComparison}
	\centering
	\begin{tabular}{l|c|c|c|c}
		\hline
		\hline
		Tool                &Extraction&Navigation&Query Language&Built-in Analyses            \\
		\hline
		JavaTA              &$\surd$     &$\times$    &$\surd$             &$\surd$            \\
		Opium               &$\surd$     &$\surd$     &$\surd$             &$\surd$            \\
		Coca                &$\surd$     &$\surd$     &$\surd$             &$\times$           \\
  	QBD									&$\surd$     &$\times$    &$\surd$             &$\times$           \\
		PQL                 &$\surd$     &$\times$    &$\surd$             &$\times$           \\
		PTQL                &$\surd$     &$\times$    &$\surd$             &$\times$           \\
		Hy$^+$              &$\surd$     &$\times$    &$\surd$             &$\times$           \\
		JIVE                &$\surd$     &$\surd$     &$\times$            &$\times$           \\
		ODB                 &$\surd$     &$\surd$     &$\times$            &$\times$           \\
		WhyLine             &$\surd$     &$\times$    &$\times$            &$\surd$            \\
		\hline
	\end{tabular}
\end{table}
%
\section{Conclusions and Future Work}
\label{conclusions}
We believe that our proposed logic programming approach is a simple and effective method for debugging object oriented programs. The key to our approach representing the execution history as a logic database, and employing logic queries to answer questions about previous execution states. Our proposed query catalog is based upon an extensive study of errors in object oriented programs \cite{hzgirgis:b}.  

Work is still in progress on JavaTA. Currently we are working on a programmable tool interface to JavaTA features. We are applying our technique to larger programs, in order to gain a better understanding of the methodology and its potential limitations.  We plan to make the JavaTA available as a plug-in for Eclipse.  We are also exploring the performance characteristics in terms of both the space and the time needed for various types of queries.  We are also interested in quantifying the overhead of extracting the program trace.
%
\subsubsection*{Acknowledgments.} The authors are thankful to the anonymous referees for their comments and suggestions especially bringing to our attention closely related work. The JavaTA logger used some code from the JDPA package from the JIVE system developed by Paul Gestwicki at the University at Buffalo.

\end{document}